# Material and Process Tolerant High Efficiency Solar Cells with Dynamic Recovery of Performance

Nithin Chatterji[#], Swasti Bhatia, Anil Kumar, Aldrin Antony, and Pradeep R. Nair[#]
Dept of Electrical Engineering, Indian Institute of Technology Bombay, Mumbai India
([#]corresponding authors: nithinchatterji@gmail.com, prnair@ee.iitb.ac.in )

*Abstract* – Low cost, highly efficient, and stable solar cells demand low temperature processing, less stringent criteria on materials, and possibility of dynamic recovery from long term degradation – a combination of features unachievable from the perspectives of current c-Si technology. To this end, here we propose a novel solar cell architecture with an additional control gate. Our simulation results indicate that the proposed device can achieve excellent efficiency even if the back-surface passivation is sub-optimal – thus allowing exploration of a wide variety of materials and low temperature fabrication processes. Importantly, such solar cells can dynamically offset efficiency loss due to elevated temperature and interface degradation associated with long term field operation and hence could be of broad interest to the PV community.

## I. INTRODUCTION

Conversion of solar energy to electricity is typically achieved through semiconductors with photovoltaic properties[1], where absorption of photons results in creation and separation of electron-hole pairs. Power conversion efficiency of such solar cells is critically influenced by the rate of recombination of photo-generated electron hole pairs[2]. Not surprisingly, modern high efficiency c-Si solar cells[3] employ diverse surface passivation techniques, which require extensive experimental calibration, to reduce the recombination of carriers at interfaces between Si and various other materials used[4], [5]. In addition, long term field operation leads to degradation of such material interfaces which could result in a time dependent reduction in the power output from solar panels[6]–[9]. So, there is a need for a scheme which could address the ill effects of interface recombination. Ideally such a scheme should be (a) process and material tolerant, (b) can be dynamic such that long term degradation of material interfaces, even if it happens, can be countered such that the power output from panels is not compromised, and (c) it should be easy to implement.

We address the above-mentioned issues by the concept of a novel "Gated Solar Cell". Here with the help of an additional metal contact, we show that the proposed solar cell can achieve high efficiency even in the presence of significant interface trap states, which could be present either due to sub-optimal initial processing or generated as a result of long term field operation. Below we first quantify the efficiency loss due to interface degradation and then illustrate how our proposed solar cell architecture achieves dynamic recovery of such efficiency losses – even at elevated temperatures.

## II. EFFICIENCY LOSS DUE TO INTERFACE DEGRADATION

Schematic of a typical c-Si based solar cell is shown in Fig. 1a. Here electron hole pairs are generated in the c-Si substrate by the solar irradiance. The emitter could be highly doped Si or electron selective layers (ESL) with appropriate band alignment to collect electrons at the top electrode[5]. The localized bottom electrode contacts the c-Si substrate through either heavily doped Si or hole selective layers (HSL) with appropriate band offset properties[3], [5]. The back surface passivation layer (BSPL) is to reduce the efficiency loss due to carrier recombination at back interface. To explore the influence of the same, we performed detailed numerical simulations (see Appendix A-B for details).

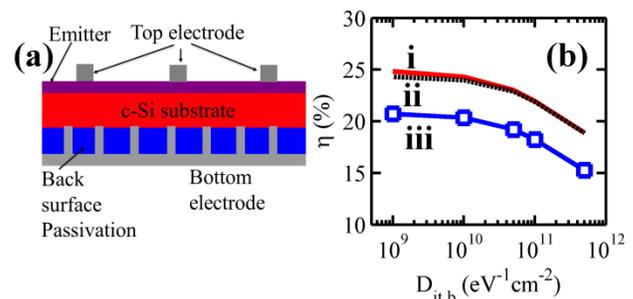

**Fig. 1**: Effect of interface degradation on traditional c-Si solar cells. (a) Schematic of a c-Si solar cell with back surface passivation. (b) Simulation results on the influence of $D_{it,f}$, $D_{it,b}$ and temperature on efficiency. Here, the cases considered are i: T = 300K, $D_{it,f}$ = 0, ii: T = 300K, $D_{it,f}$ = $10^{11}$, iii: T = 343K, $D_{it,f}$ = $10^{11}$ (units of $D_{it}$: $cm^{-2}eV^{-1}$).

Fig. 1b shows the efficiency variation as a function of both front and back surface quality in terms of the interface trap density ($D_{it}$) denoted by $D_{it,f}$ for the front surface (i.e., ESL/Si) and $D_{it,b}$ for the back surface (i.e., Si/BSPL), respectively. The results indicate that – (a) for excellent surface passivation (i.e., small $D_{it}$), the efficiency (η) is very high and is limited by bulk recombination effects. (b) An increase in either $D_{it,f}$ or $D_{it,b}$ results in efficiency degradation. (c) Surprisingly, for similar value of interface trap density (say, $10^{11} cm^{-2}eV^{-1}$), $D_{it,b}$ degrades efficiency more than $D_{it,f}$ - in spite of the fact that peak photo-carrier generation happens near the front surface. This relative insensitiveness to $D_{it,f}$ is due to the strong field effect passivation associated with increased band bending at the ESL/Si heterojunction, as reported recently [10]. Fig. 1b also indicates that efficiency degradation worsens at elevated temperatures expected under field conditions (due to an increase in carrier recombination which depends on intrinsic carrier density [2], [11]). All these indicate that schemes tolerant to interface passivation quality are highly desirable from multiple perspectives.

### III. DYNAMIC RECOVERY OF EFFICIENCY

The efficiency degradation due to poor quality of back surface passivation can be effectively addressed through our proposed structure shown in Fig. 2a. Here, with the help of an additional "Control Gate" (shown in green color which are all interconnected through a patterned grid and held at same potential, $V_g$) the carrier density at this interface could be modulated which results in reduced carrier recombination (see Appendix C) and hence improved efficiency. We note that similar field effect phenomena improve the passivation quality of dielectrics like $Al_2O_3, SiO_2$, etc. [12], [13]. Indeed, the numerical simulation results (Fig. 2b) indicate that efficiency loss due to carrier recombination at back surface could be almost entirely recovered through appropriate control gate bias – proof of concept for the proposed architecture. As expected, the control gate allows biasing the semiconductor in all three regimes of operation – accumulation ($V_g < -0.3V$), depletion or inversion ($V_g > -0.3V$, also see Appendix B). Further, the efficiency recovery is significant even at elevated temperatures expected during field operation (see curve iii) – a feature of immense importance for commercial applications.

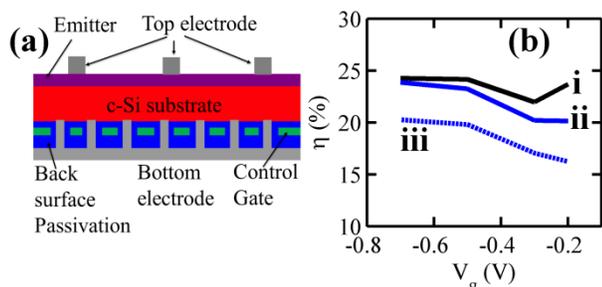

**Fig. 2**: Efficiency recovery through proposed gated solar cell architecture. (a) Schematic of proposed solar cell with additional control gate. (b) Simulation results showing efficiency recovery through control gate in the presence of degradation at the interface between c-Si and back surface passivation ($D_{it}$ units: $cm^{-2}eV^{-1}$). Here the conditions are, i: $D_{it,b} = 10^{11}$, T = 300K, ii: $D_{it,b} = 5 \times 10^{11}$, T=300K, and iii: $D_{it,b} = 5 \times 10^{11}$, T = 343K, with $D_{it,f} = 10^{11}$ in all cases.

### IV. DISCUSSIONS

Our proposal could be of potential interest to the community on multiple accounts – from device to system level. From the perspectives of device fabrication, it is evident that: (a) with the control gate, excellent performance can be achieved even when the interface quality is poor. As a result, many materials could be explored for BSPL with less stringent criteria on the quality of interface passivation. (b) Our scheme results in performance improvement over a wide range of bias conditions – not just in inversion but also in accumulation conditions. (c) The proposed scheme is applicable for solar cell architectures with homojunctions (PERC[3], PERL[14], etc.) and is easily adaptable to heterojunction based carrier selective architectures including IBC (Interdigitated Back Contact[15]) solar cells. (d) There could be significant advantages on the thermal budget, especially when used along with carrier selective architectures, as high temperature anneal steps could be avoided in our scheme – thus, potentially, reducing the cost. In addition, (e) the proposed scheme could provide additional flexibility in the passivation quality vs. reflectivity trade-off involved with nitride/carbide dielectric layers[16] and hence could also be appealing for thin c-Si solar cells[17].

In addition, there are some interesting aspects at the system level: (i) The control gate is effective even if the interface quality degrades with time – thus enabling prospects for dynamic recovery of efficiency. Such a scheme could alleviate any time dependent interface degradation component that might contribute towards LeTID (Light and elevated Temperature Induced Degradation [18]). (ii) Our results indicate that efficiency recovery is indeed possible even at elevated temperatures expected during field operation. It will be interesting to explore the possible gains accrued over long term field operation. (iii) Note that as the control gate voltage appears across an insulator (i.e., BSPL), the electrical load associated is capacitive – much like a MOSFET[19]. Accordingly, there is no dynamic power loss associated with the control gate during field operation. Interestingly, simple electronic circuitry could enable the control gate voltage to be tapped from the module output itself – with no significant power loss during steady state operation. Design of such a scheme so as to optimize the overall power output, associated cost calculations, and the eventual impact on Levelized cost of electricity (LCOE) could be an interesting challenge at system level.

### V. CONCLUSIONS

To summarize, current industry standard solar cells require extensive experimental research to improve the quality of material interfaces to reduce the photo-carrier recombination. This introduces severe constraints in terms of the thermal budget and the materials that could be used for surface passivation. A scheme which could dynamically reduce interface recombination – regardless of the type of material used, the initial quality of the interface and how it degrades with time – is highly desirable, if possible. Our proposal is a new architecture for solar cells which requires significantly less process optimization and allows much more, potentially cheaper, alternatives for surface passivation. Interestingly, the proposed scheme can reduce power loss at elevated temperature under field conditions. While experimental efforts to demonstrate the proof of concept are on-going, the proposed solar cell architecture has distinct advantages which should be of broad interest to the PV community from multiple aspects ranging from lab scale efforts to commercial production.

**Appendix A. Simulation Methodology**: Numerical simulation of the structure shown in Fig. 1a or Fig. 2a, through self-consistent solution of Poisson and drift-diffusion equations for carriers [2], [20], is challenging due to multiple accounts: in addition to the scale of physical dimensions involved, there are severe numerical convergence issues due to the presence of localized top contacts, interface recombination, and position dependent photo-carrier generation rate. Consequently, the results presented here were obtained using the structure shown in Fig. 3 with appropriate photo-carrier generation rate. Accordingly, the numerical simulations using Fig. 3 differs from that of Fig. 2a only in terms of the resistive loss due to lateral transport of photo-generated carriers through ESL to the localized top contacts, and nothing else. Indeed, such resistive losses were found to be negligible as determined through comparable set of simulations (between the structures shown in Figs. 1a or 2a with Fig. 3. Results not shared here).

The position dependent photo-carrier generation rate in Si was obtained with 1Sun illumination as input spectra and wavelength dependent absorption coefficient for Si[21], [22] (which results in $J_{sc} \sim 40 mAcm^{-2}$). As expected, most of the carrier generation happens very close to ESL/Si junction. Heavily doped ESL (N-type) and HSL (P-type) are used to selectively extract electrons and holes, respectively. Trap states with uniform density are considered both at ESL/Si and at Si/BSPL interfaces, with capture cross section of $10^{-16} cm^{-2}$ (for both electrons and holes [23], [24]). Note that the structure shown in Fig. 3, in the absence and presence of control gate, was used to obtain the results of Fig. 1b and Fig. 2b, respectively. We remark that the results shown in Fig. 2b could depend on the work function difference between control gate and c-Si (assumed as $-0.34eV$ in this study) and the capacitance of BSPL layer[19]. Further, the temperature dependent simulations were done as per prior literature [11]. The various material parameters assumed are provided in Appendix B.

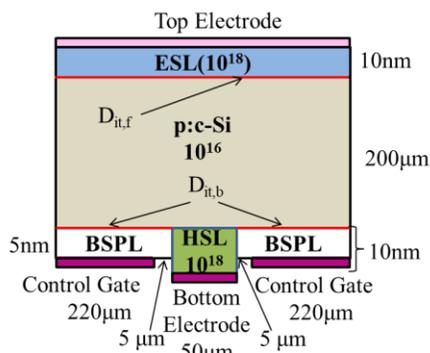

Fig. 3: Structure of the device used in numerical simulations.

**Appendix B. Parameters used in Simulations:** The band gaps of Si, ESL, and HSL are 1.12eV, 3.4eV, 2.12eV respectively. The electron affinities for Si, ESL, HSL are 4.05eV, 4.05eV, 3.05eV respectively. For c-Si, SRH lifetime is 1ms, the radiative recombination coefficient is $1.1 \times 10^{-14} cm^3 s^{-1}$ and the Auger coefficients for electrons and holes are $1 \times 10^{-31}$ and $0.79 \times 10^{-31}$ $cm^6 s^{-1}$, respectively. BPSL properties are similar to $SiO_2$ and is assumed to be charge free. Any charge in BSPL could contribute to a flat-band voltage change and can be easily incorporated in our simulation framework[20].

The temperature dependent simulations are based on the methodology described in reference [11]. Briefly, the parameters which have significant temperature dependences are: (i) band gap [25] specified as $E_g(T) = E_g(0) - \frac{\alpha T^2}{T+\beta}$, where $E_g(T)$ is the band gap at T and $E_g(0)$ is the band gap at 0K, with $\alpha = 4.73 \times 10^{-4} eV K^{-1}$, and $\beta = 636K$. (ii) carrier mobility [20] modeled as $\mu_{constant} = \mu_L \left(\frac{T}{300K}\right)^{-\varsigma}$. The values of $\mu_L, \varsigma$, and the rest of the parameters are the same as used in previous publications [10], [11].

**Appendix C. Control gate induced field effect passivation**: Fig. 4a compares the IV characteristics of the device shown in Fig. 1a with that of Fig. 2a. The use of control gate results in the reduction of carrier recombination at back interface and hence an improvement in $V_{oc}$ and the efficiency. Part (b) shows the influence of control gate on the average carrier recombination rate at the Si/BSPL interface ($R_{interface}$, i.e., averaged over the entire area) of device shown in Fig. 2a (under 1Sun illumination and at $V = V_{mpp} \approx 0.57V$ as indicated in part (a)). The control gate bias results in modulation of carrier density at Si/BSPL interface and the carrier recombination rate peaks at $V_g = -0.3V$ (not at 0V because of the workfunction difference between control gate and Si). Using the gate bias, the interface recombination component can be reduced (known as field effect passivation) which results in an increase in efficiency. In PV applications, field effect passivation is usually achieved through fixed charges, if any, present in dielectric layers [12]. Here, the control gate is a far more flexible scheme which could enable additional functionalities.

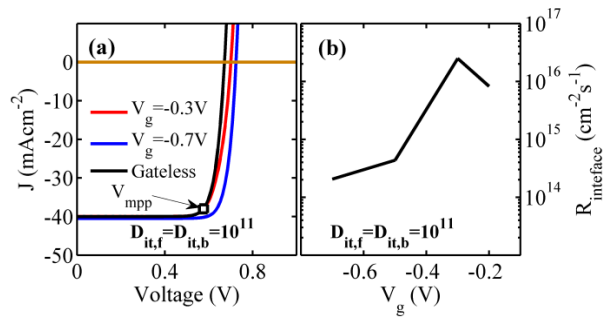

Fig. 4: Control gate induced field passivation of Si/BSPL interface. Part (a) compares the IV for gated solar cell for different $V_g$ with the device in Fig. 1a. Part (b) shows the $V_g$ dependent $R_{interface}$ at Si/BSPL interface at an applied forward bias of 0.57 V (indicated as $V_{mpp}$ in part (a)).

**Acknowledgements:** The authors acknowledge Prof. K L. Narasimhan and Prof. B. M. Arora, both from EE, IIT Bombay, for engaging discussions. The authors also acknowledge IIT Bombay Nanofabrication Facility (IITBNF) and National Center for Photovoltaic Research and Education (NCPRE), IIT Bombay for computational facilities. PRN acknowledges Visvesvaraya Young Faculty Research Fellowship, DST India.

**Updated version of this manuscript, with additional details is accepted for publication in IEEE Transactions on Electron Devices.**

**Updated version of this manuscript, with additional details is accepted for publication in IEEE Transactions on Electron Devices.**